\documentclass[%
 reprint,
superscriptaddress,
 amsmath,amssymb,
 aps,
 prl,
]{revtex4-1}

\usepackage{graphicx}
\usepackage[space]{grffile}
\usepackage{latexsym}
\usepackage{textcomp}
\usepackage{longtable}
\usepackage{multirow,booktabs}
\usepackage{amsfonts,amsmath,amssymb}
\usepackage{natbib}
\usepackage{url}
\usepackage{hyperref}
\hypersetup{colorlinks=false,pdfborder={0 0 0}}
\newif\iflatexml\latexmlfalse
\DeclareGraphicsExtensions{.pdf,.PDF,.png,.PNG,.jpg,.JPG,.jpeg,.JPEG}

\usepackage[utf8]{inputenc}
\usepackage[english]{babel}
\usepackage{color}

\begin{document}

\title{Bootstrapping to the Molecular Frame with Time-domain Photoionization Interferometry}

\author{Claude Marceau}
\affiliation{Joint Attosecond Science Laboratory, National Research Council of
Canada and University of Ottawa, 100 Sussex Drive, Ottawa, K1A 0R6,
Canada}

\author{Varun Makhija}
\affiliation{Department of Physics, University of Ottawa, 150 Louis Pasteur, Ottawa, ON K1N 6N5, Canada}

\author{Dominique Platzer}
\affiliation{Joint Attosecond Science Laboratory, National Research Council of
Canada and University of Ottawa, 100 Sussex Drive, Ottawa, K1A 0R6,
Canada}

\author{A. Yu. Naumov}
\affiliation{Joint Attosecond Science Laboratory, National Research Council of
Canada and University of Ottawa, 100 Sussex Drive, Ottawa, K1A 0R6,
Canada}

\author{P. B. Corkum}
\affiliation{Joint Attosecond Science Laboratory, National Research Council of
Canada and University of Ottawa, 100 Sussex Drive, Ottawa, K1A 0R6,
Canada}

\author{Albert Stolow}
\affiliation{Department of Physics, University of Ottawa, 150 Louis Pasteur, Ottawa, ON K1N 6N5, Canada}
\affiliation{Department of Chemistry, University of Ottawa, 10 Marie Curies, Ottawa, ON K1N 6N6, Canada}
\affiliation{National Research Council of Canada, 100 Sussex Drive, Ottawa, K1A 0R6, Canada}

\author{David Villeneuve}
\affiliation{Joint Attosecond Science Laboratory, National Research Council of
Canada and University of Ottawa, 100 Sussex Drive, Ottawa, K1A 0R6,
Canada}

\author{Paul Hockett}
\email{paul.hockett@nrc.ca}
\affiliation{National Research Council of Canada, 100 Sussex Drive, Ottawa, K1A
0R6, Canada}

\begin{abstract}
Photoionization of molecular species is, essentially, a multi-path interferometer with both experimentally controllable and intrinsic molecular characteristics. In this work, XUV photoionization of impulsively aligned molecular targets ($N_2$) is used to provide a time-domain route to ``complete" photoionization experiments, in which the rotational wavepacket controls the geometric part of the photoionization interferometer. The data obtained is sufficient to determine the magnitudes and phases of the ionization matrix elements for all observed channels, and to reconstruct molecular frame interferograms from lab frame measurements. In principle this methodology provides a time-domain route to complete photoionization experiments, and the molecular frame, which is generally applicable to any molecule (no prerequisites), for all energies and ionization channels.
% In principle this methodology is generally applicable, since there are no molecule-specific prerequisites, for all energies and ionization channels.
% This methodology provides a time-domain route to complete photoionization experiments, and the molecular frame, which is generally applicable to any molecule.
% and provides a time-domain route to complete experiments which is generally applicable.
%Although complicated, with high information-content observables the interferometric nature of the process allows for the magnitudes and phases of the (intrinsic) ionization matrix elements to be determined - this is the aim of ``complete” photoionization studies. 
%In this work, photoionization of impulsively aligned molecular targets ($N_2$) to provide a time-domain route to complete experiments, in which the rotational wavepacket controls the geometric part of the photoionization interferometer, and provides a time-domain route to complete experiments which is generally applicable.
%; specifically, a double-pulse scheme is used to create a rotational wavepacket in an ensemble of $N_{2}$ molecules, which are later ionized by a pulse train of high harmonics. From this data, sets of ionization matrix elements can be extracted for various ionization pathways via fitting of the time-domain data within a canonical photoionization formalism. This methodology, in which the rotational wavepacket controls the geometric part of the photoionization interferometer, provides a time-domain route to complete experiments which is generally applicable.%
\end{abstract}%

\maketitle

% Full author list (revtex formatted, for export):
% Claude Marceau, Varun Makhija, Dominique Platzer, A. Yu. Naumov, P. B. Corkum, Albert Stolow  D. M. Villeneuve and Paul Hockett

Photoionization is an interferometric process, in which the final observable
results from a coherent sum over multiple quantum paths to a set of
final continuum photoelectron states $|k,l,m\rangle$ \cite{Cohen_1966, Dill_1976}.
Interferences between these components are manifest in the observable
energy spectra and photoelectron angular distributions (PADs) \cite{Cooper1969,Dill_1976,Reid_2003},
the latter of which can be considered as a particularly high information
content observable, extremely sensitive to the phases of the partial
waves $|l,m\rangle$ \cite{Reid_2003, Ramakrishna_2012}; consequently, PADs have been
investigated in a large range of control and metrology scenarios \cite{Reid_2012,Hockett_2015b} .
In the context of phase-sensitive metrology, the goal is to obtain
the full set of complex photoionization matrix elements, hence characterise the photoelectron wavefunction, by analysis of sets of PAD measurements - this is a ``complete'' photoionization experiment \cite{Klar_1982,Becker_1998}.

In the molecular case, the number of final $|l,m\rangle$ states is
typically large, and obtaining a sufficient dataset for a complete
experiment remains a challenge. In the energy domain, a number of different
schemes have been demonstrated in both the laboratory (LF) and molecular
frames (MF) \cite{Reid_1992,Shigemasa_1995,Gessner_2002,Lucchese_2002,Hockett_2009,Tang_2010,Hockett_2014}. The common theme to all these measurements is some form of control over the experimental contributions to the
photoionization interferometer (e.g. rotational state, polarization geometry), 
to which the intrinsic molecular contributions
remain invariant. The difficulties in such cases are the ability to
obtain a sufficiently large dataset, and molecular specificity in
the methodologies, i.e. prerequisites such as low density
of states \cite{Hockett_2009}, resonances \cite{Tang_2010} or dissociative channels \cite{Shigemasa_1995,Gessner_2002,Lucchese_2002}.

In the time-domain, rotational wavepackets can be utilized to control the geometric part of the interferometer.
In this case, a high degree of spatio-temporal control of the axis
distribution (alignment) of the ionizing molecular ensemble in the
LF can be obtained via preparation of a broad rotational wavepacket. Although this idea is conceptually obvious, the theory is complex; it has
 been elucidated by multiple authors (e.g. refs. \cite{Underwood_2000,Seideman_2002,Suzuki_2006,Suzuki_2007,Ramakrishna_2012,Hockett_2015}), but - to date - there have been no experimental demonstrations beyond the limiting case of a narrow wavepacket prepared
via resonant excitation \cite{Suzuki_2006,Tang_2010,Suzuki_2012} (c.f. rotational coherence spectroscopy \cite{Felker_1987,Felker_1992}), and an exploration in the related case of high-harmonic spectroscopy from an impulsively aligned sample \cite{Lock_2012}. 
While experimental methods for
preparing rotational wavepackets, and measuring PADs, are relatively
well established \cite{Eppink_1997,Suzuki_2006,Reid_2012,Rouz_e_2012}, the analysis of such experiments remains
challenging, since both the rotational wavepacket and the photoionization
dynamics must now be fully characterised. However, the benefits are
significant - the time-domain methodology is, in principle, \emph{completely
general with no molecule-specific prerequisites}; additionally the
use of high harmonics for ionization provides channel and energy multiplexing
in each time-domain measurement, resulting in an extremely high information
content metrology \cite{Suzuki_2005,Ramakrishna_2012,Rouz_e_2012}. Furthermore,
if the determination of the photoionization matrix elements is of
sufficient fidelity, one can reconstruct the MF interferograms \emph{without
the necessity for direct MF measurements}. In general, such measurements are desirable for detailed understanding of molecular processes \cite{Lucchese_2012}.

% In this work, the utility of such a methodology is demonstrated: 
Here we present a general time-domain approach to complete photoionization experiments in molecules. 
Experimentally, a double-pulse impulsive alignment scheme is used to create a broad
rotational wavepacket in N$_2$~\cite{Bisgaard_2004,Cryan_2009,Ren_2013}. High harmonics of a 267 nm driving field are
used to ionize the aligned ensemble, and velocity-map imaging (VMI)
provides energy, angle and time-resolved photoelectron
interferograms. A bootstrapping methodology is employed to analyse
the data, in which (1) the prepared rotational wavepacket is characterised
without knowledge of the intrinsic molecular photoionization dynamics \cite{Makhija2016}
and (2) these results feed into a protocol for the determination of the channel-specific ionization matrix elements.
% via a fitting protocol which incorporates the axis distributions determined
% in the preceding step. 
As a test of the results so obtained, the MF
interferograms (as a function of polarization geometry) are reconstructed and compared with \emph{ab initio}
calculations.  The key results are presented herein; an extended presentation and discussion, and numerical data, can be found in the \href{https://dx.doi.org/10.6084/m9.figshare.4480349.v1}{Supplementary Material (SM)} \cite{SM}.

% \textcolor{red}{{[}That might be a little too detailed, but this is what I have so far{]}}
% \textcolor{red}{{[}VMI data is shown in fig. 1{]}}

\begin{figure}[h!]
\begin{center}
\includegraphics[width=1\columnwidth]{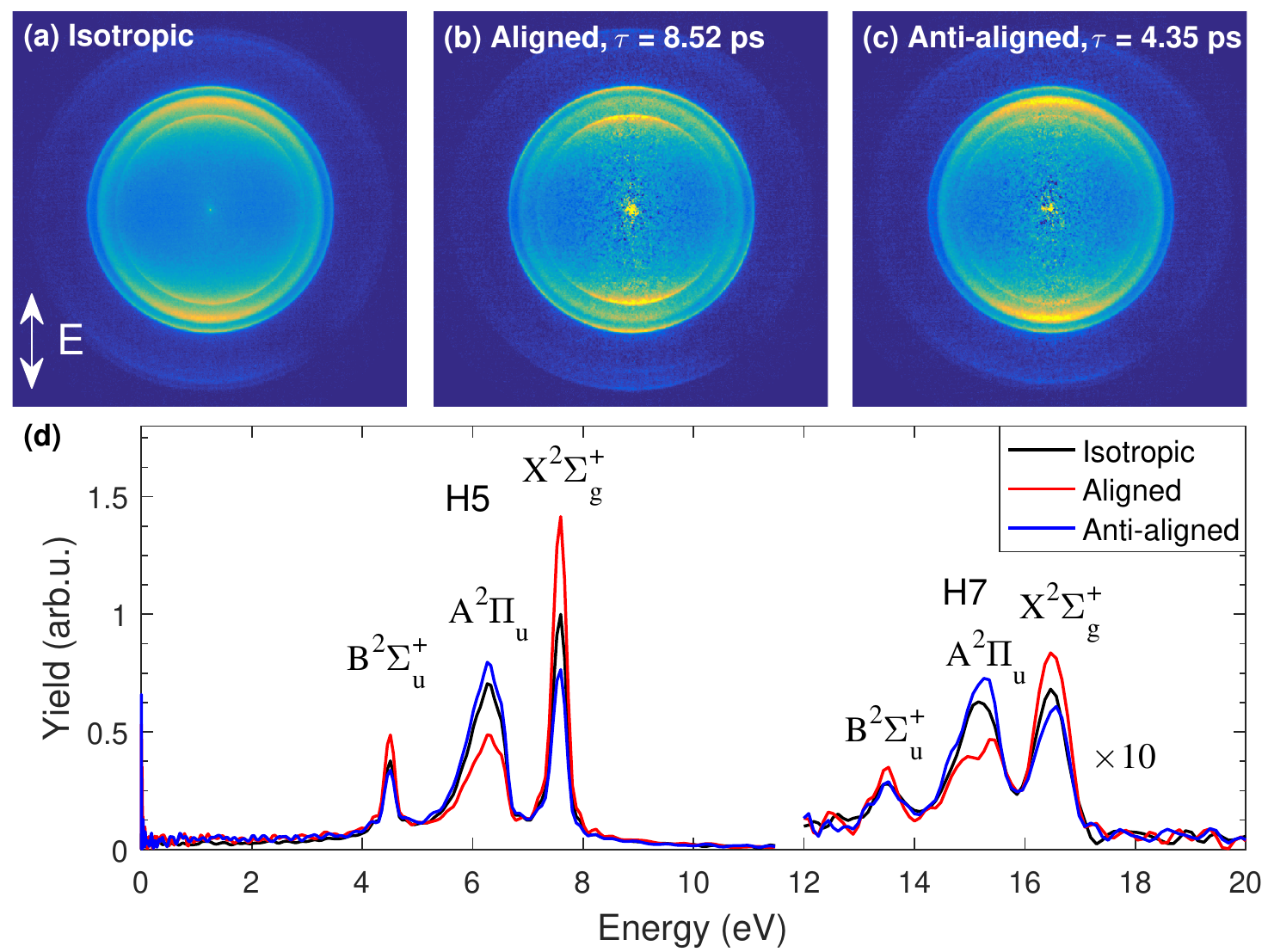}
\caption{{\label{fig:VMIresults}VMI photoelectron images from (a) isotropic (no alignment pulses) (b) aligned (c) anti-aligned nitrogen
molecules. The polarization of both alignment pulses and the XUV pulse are linear as indicated in (a). (d) The electron yields, labels provide cationic state assignments for the observed features.%
}}
\end{center}
\end{figure}

\textit{Experiment}: laser pulses were generated by a Ti:sapphire amplifier (100~Hz, 11~mJ, 800~nm, 50~fs). Approximately 3.5~mJ was used to generate the third harmonic (267~nm, $\sim$0.05~mJ, $\sim$100~fs) via double and sum mixing stages in $\beta$-BBO. 
% were split into two by a beam splitter. The reflected part (3.5~mJ) was used to generate the third harmonic (267~nm, $\sim$0.05~mJ, $\sim$100~fs) via double and sum mixing stages in $\beta$-BBO.
% was first frequency doubled in a 0.5 mm-thick $\beta$-BBO. After a 0.75 mm-thick calcite plate (delay compensator) and a half-wave plate to rotate the fundamental polarization, the third harmonic (267 nm) was generated by sum frequency of both the fundamental and its second harmonic inside a  0.15 mm-thick $\beta$-BBO. 
The output was filtered by dichroic mirrors and then sent into a vacuum chamber where high harmonics were generated from a pulsed gas jet of argon. %Laser parameters for high harmonic generation at 267 nm were about 0.05 mJ and 100 fs. 
Harmonics were then sent through a 200 nm-thick aluminum foil to filter out the driving laser light; the final spectrum was dominated by harmonics 5 (H5, $h\nu$=23.3~eV) and 7 (H7, $h\nu$=32.6~eV). %Harmonics 5 and 7 could be seen in the photoelectron spectra.
The remainder of the initial 800~nm beam (7.5~mJ) was sent to a Michelson interferometer; the two replicas of the pulse generated formed alignment pulses. % A downsizing telescope was used to fine tune the focal region and overlap of these two alignment pulses and the harmonic beam. %Focal lengths were +125 mm and -75 mm
The alignment and filtered harmonic beams were recombined (on a holey mirror) and focussed into the interaction region of a VMI spectrometer \cite{Eppink_1997}.
%(2~mm diameter hole). The resulting collinear beams were focused by a gold-coated toroidal mirror (F=400 mm, AOI 85 deg) into a velocity map imaging (VMI) spectrometer, and intersected a gas jet in the interaction region of the spectrometer. 
Spatial and temporal superposition of the alignment pulses and the harmonic pulses was achieved by maximizing the AC Stark shifts in the photoelectron spectrum of argon. %Additional proof of the good spatial superposition comes from cation imaging in the VMI, which comes from the same 50 $\mu$m width for both the harmonic and the alignment pulses.
The estimated laser parameters in the interaction region for each alignment pulse are 0.5~mJ, 100 fs at focus, % (slight positive chirp and some self-phase modulation), 
resulting in a peak intensity $I$~=~20~TW/cm$^{2}$. All beams were linearly polarized, and the data reported herein was obtained for a parallel polarization geometry (Fig. \ref{fig:VMIresults}(a)). The delay between the two pump pulses was experimentally optimized (for maximal alignment) at 3.76~ps, near the rising edge of the half revival induced by the first pulse~\cite{Ren_2013}.

Figure \ref{fig:VMIresults} presents three example photoelectron images from molecular nitrogen (99.999\% purity, 4.8~bar backing pressure, 250~$\mu m$ nozzle) and corresponding photoelectron spectra. % for (a) the isotropic case (no alignment pulses), (b) the aligned case and (c) the anti-aligned case, and the corresponding photoelectron spectra.  %The alignment pulses create a significant ATI spectrum by themselves, which was subtracted from the images before data processing. 
The complete dataset consists of 150 temporal steps in 67~fs increments from $t=$-0.5~ps to +9.5~ps, where $t$=0 is defined as the peak intensity of the second alignment pulse. On average, about 30000 laser shots were accumulated at each delay. %To minimize laser and gas density fluctuation effects, the delay was scanned several times. % taken smaller samples on many occasions. 
Additional signals that monitored gas density in the VMI, harmonic source brightness and background contributions (ATI from the alignment pulses, scattered light, residual background gas signals) were taken together with the data, providing calibration and background measurements. % \textcolor{red}{[further details in SM?]}
%providing a calibration for slow drifts over the acquisition time, and background signals for subtraction. % and subtracted taking into account slow drifts over the acquisition time.  
% Data points around $\tau$=0 were discarded because sum frequency processes complicate the photoelectron spectrum when there is significant overlap between pump and probe. 
% Data analysis consisted of inverting each background-subtracted VMI image using the pBaseX algorithm \cite{Garcia_2004}; the resulting Legendre polynomial expansion coefficients were averaged over each photoelectron band (Fig. \ref{fig:VMIresults}) [integrate with next para],
%, corresponding to the three final electronic states of N$_{2}^{+}$ populated via ionization from H5 and H7, namely  $X^{2} \Sigma_{g}^{+}$, $A^{2} \Pi_{u}$ and  $B^{2} \Sigma_{u}^{+}$ 
% hereafter denoted as the $X$, $A$ and $B$ channels. Time-domain results are shown in Figs. \ref{fig:ADM_fits}, \ref{fig:BLMfits} for the $X$ channel of the fifth harmonic.

\begin{figure}[h!]
\begin{center}
\includegraphics[width=1\columnwidth]{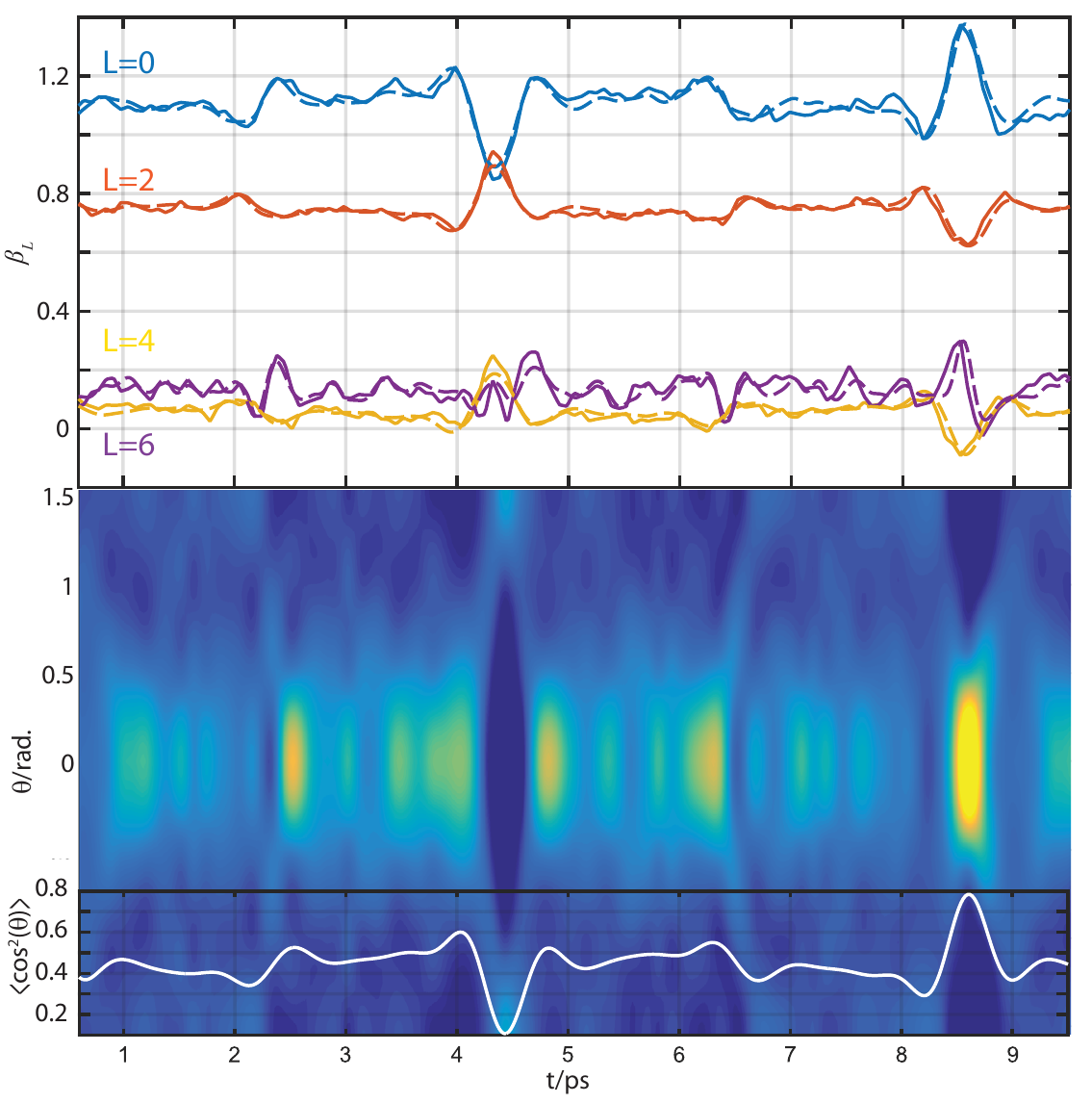}
\caption{{\label{fig:ADM_fits}Molecular alignment determination. (Top) Experiment (solid lines) and fit results (dashed) to eqn. \ref{eq:model}, for $L=0,2,4,6$ (Legendre polynomial expansion). (Bottom) Calculated axis distribution $P(\theta,t)$ determined from the fit ($I$~=~20~TW/cm$^2$, $T_{rot}$~=~15~K), inset shows $\langle \cos^2(\theta,t)\rangle$.  The temporal axis is defined by the second pulse (t~=~0), and shows the 1/2 and full revivals at the expected delays of 4.2 and 8.4~ps respectively. % The data shown and used for the fit is a delay scan after the second pulse, showing 1/2 and full revivals at the expected delays of 4.2 and 8.4~ps respectively. %
}}
\end{center}
\end{figure}

\begin{figure}[h!]
\begin{center}
\includegraphics[width=1\columnwidth]{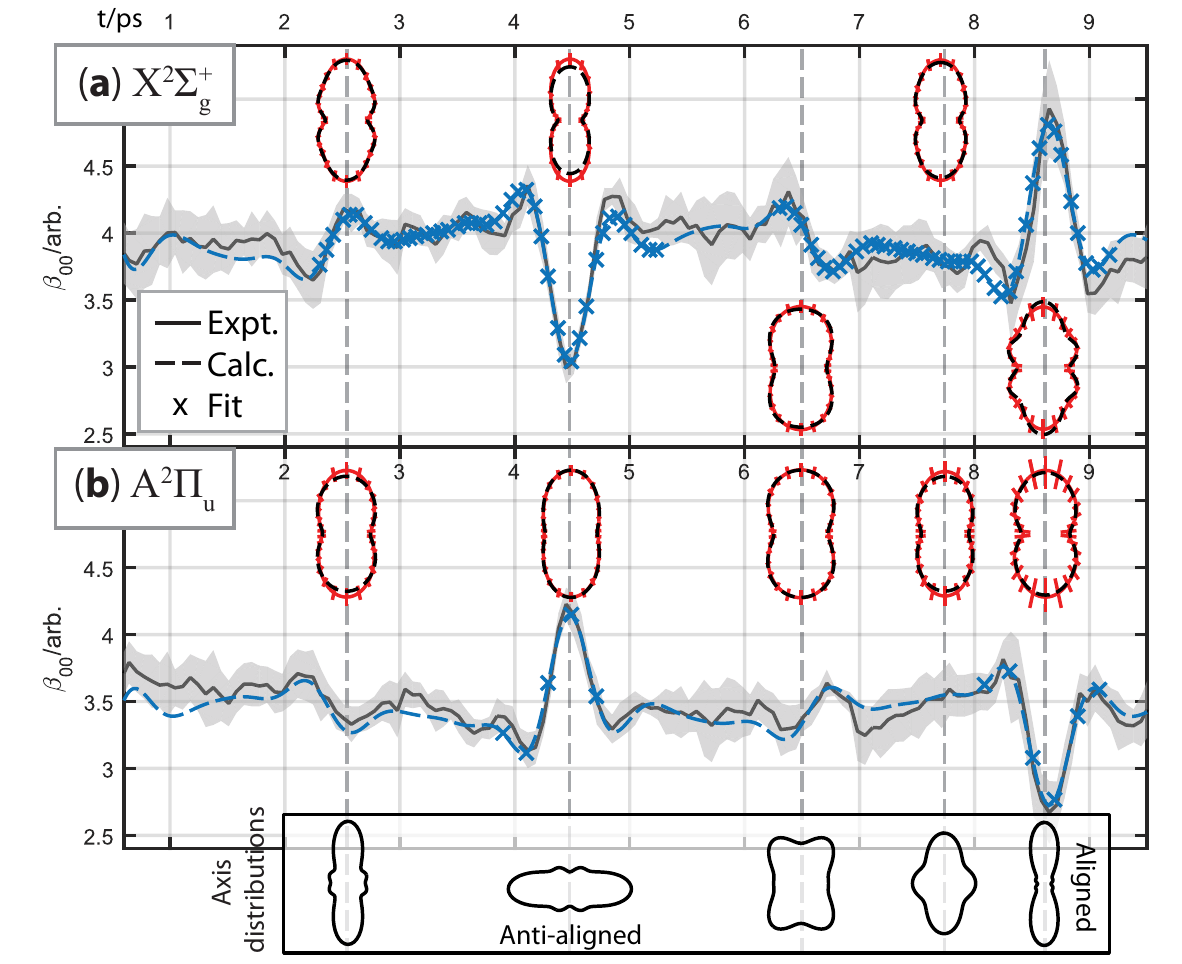}
\caption{{\label{fig:BLMfits}Fitting of the the time-series data. Main panels show the calculation results from Eqn. \ref{eq:beta} for the $X$ and $A$-channel yields, $\beta_{00}(t)$, `x' marks points included in the fit; full $\beta_{L,M}(t)$ plots are presented in the SM. % and the grey envelopes indicate the experimental uncertainties, derived as the standard-deviation of repeat measurements. 
Examples of the full angle-resolved interferograms $S(\theta,t)$, experimental results inc. error bars and calculated, are also shown, and corresponding axis distributions $P(\theta,t)$ (bottom). % at specific times, experimental results with error bars and calculated. 
% (Inset) Axis distributions $P(\theta,t)$ corresponding to the $S(\theta,t)$ plots. 
% (Top) Full $\beta_{LM}(t)$ results for the  $X$-channel. NOW REMOVED%
}}
\end{center}
\end{figure}

\textit{Theory}: The observed angular distributions are most generally expressed as an expansion in spherical harmonics:

\begin{equation}
S(\theta,\phi,t)=\sum_{L,M}\beta_{L,M}(t)Y_{L,M}(\theta,\phi)
\label{eq:St}
\end{equation}
where $\beta_{L,M}(t)$ are the expansion coefficients. For cylindrically symmetric cases, as for all LF quantities discussed herein, $M=0$, hence the angular dependence is reduced to a function of $\theta$, and the angle $\phi$ is redundant. % (i.e. functionally equivalent to an expansion in Legendre polynomials $P_L(\cos (\theta ))$). 
The LF symmetry also restricts terms to even-$L$ only.  The experimental data analysis consisted of inverting each background-subtracted VMI image using the pBaseX algorithm \cite{Garcia_2004}, providing expansion coefficients for each photoelectron image as a function of % radius, i.e. $\beta_{L,M}(r;~t)$, with $L_{\mathrm{max}}=6$ determined from the experimental images; these were averaged over each
radius, subsequently averaged over each photoelectron band (Fig. \ref{fig:VMIresults}(d)) to provide $\beta_{L,M}(t)$ for each ionization channel,
%, corresponding to the three final electronic states of N$_{2}^{+}$ populated via ionization from H5 and H7, namely  $X^{2} \Sigma_{g}^{+}$, $A^{2} \Pi_{u}$ and  $B^{2} \Sigma_{u}^{+}$ 
hereafter denoted as the $X$, $A$ and $B$ channels. Selected time-domain results are shown in Figs. \ref{fig:ADM_fits}, \ref{fig:BLMfits} for H5 $X$ channel (full presentation in SM \cite{SM}).

The $\beta_{L,M}(t)$ can further be expanded in terms of the contributing physical factors:

\begin{equation}
\beta_{L,M}(t)=\sum_{K,Q}\left(\sum_{\alpha,\alpha'}\gamma_{K,Q}^{\alpha,\alpha'}\boldsymbol{D}_{\alpha}^*\boldsymbol{D}_{\alpha'}\right)A_{K,-Q}(t)
\label{eq:beta}
\end{equation}
In this form, all of the angular momentum coupling terms
are denoted by $\gamma$, and can be defined analytically. % $E_{P,R}$ denotes the ionizing radiation field, 
$A_{K,-Q}(t)$ are the time-dependent axis distribution moments (ADMs) arising from the rotational wavepacket, and define the spatial alignment of the molecular ensemble: the full axis distribution in the LF can be described as $P(\theta,\phi,t)=\sum_{K,Q}A_{K,-Q}(t)Y_{K,Q}(\theta,\phi)$ (cf. Eqn. \ref{eq:St}). The ADMs
couple geometrically to the ionization dynamics - the intricate details of this coupling dictates the response of different partial waves to the ADMs, and different revivals in the rotational wavepacket are sensitive to different aspects of the ionization dynamics \cite{Hockett_2015}.
The $\boldsymbol{D}_{\alpha}$ are the symmetrized
ionization matrix elements (complex) to be determined. % labelled by channel $\Gamma$, as a function of the photon projection in the molecular frame $q$. 
All other required quantum numbers are denoted $\alpha$, and the coherent summation
is obtained by summing over all possible pairs of each quantum number. The full form of Eqn. \ref{eq:beta}, which explicitly shows all summations (all interfering paths) which contribute to each observable $\beta_{L,M}(t)$, is given in the SM \cite{SM}. This treatment follows the formalism of Underwood \& Reid \cite{Underwood_2000,Reid_2000,Stolow} and is applicable to single-photon ionization in the dipole limit.  % The $\boldsymbol{D}_{\alpha}$ are the matrix elements to be determined. 

Evidently, the above equation can be recast as

\begin{equation}
\beta_{L,M}(t)=\sum_{K,Q}C_{K,Q}^{L,M}A_{K,-Q}(t)
\label{eq:model}
\end{equation}
where the coefficients $C_{K,Q}^{L,M}$ contain all the terms in the brackets in Eq.~\ref{eq:beta}. Since this equation is linear in the ADMs, the measured $\beta_{L,M}(t)$ can be used to determine the ionization dynamics in this phenomenological form - such solutions of Eq.~\ref{eq:model} constitute the first part of the bootstrapping procedure \cite{Makhija2016}. 

\textit{Analysis}: The ADMs can be computed by solving the Time Dependent Schr{\"o}dinger Equation (TDSE) for a linear rigid rotor rotor in a linearly polarized non-resonant pulse. Although the $C_{K,Q}^{L,M}$ are channel dependent, the axis distribution is universal; hence fits to different $L$ and/or ionization channels can be performed as a rigourous cross-check on the results. In this case, computation of the ADMs was carried out for rotational temperatures $T_{rot}$~=~1~K to 30~K in 1~K steps and laser intensities $I$~=~10~TW/cm$^2$ to 30~TW/cm$^2$ in 2~TW/cm$^2$ steps for each pulse. The delay between pulses and their durations were kept fixed to the experimentally determined values. The calculated ADMs were stored and a linear regression to solve Eq.~\ref{eq:model} was carried out for each parameter set on the measured $\beta_{0,0}(t)$, $\beta_{2,0}(t)$ and $\beta_{4,0}(t)$ parameters for the $X$-channel. Each regression was performed independently, and $I$~=~20~TW/cm$^2$ for each pulse and $T_{rot}$~=~15~K provided the best fit in all three cases, in good agreement with the experimentally estimated values. The fits are shown in Fig.~\ref{fig:ADM_fits}, along with a comparison with $\beta_{6,0}(t)$ obtained using the same parameter set ($I$ and $T_{rot}$). Given the fidelity of the fits, the ADMs ($K_{max}=6$) obtained for these parameters are assumed to accurately describe the experimentally prepared axis distributions, and the resultant axis distribution $P(\theta,t)$ % full axis distributions $P(\theta,t)=\sum_{K,Q}A_{K,-Q}(t)Y_{K,Q}(\theta,\phi)$, where LF cylindrical symmetry dictates $Q=0$ only, % and associated $\langle \cos^2(\theta) \rangle$ are also shown 
is shown in Fig.~\ref{fig:ADM_fits}.

% For $N_{2}$, the number and type of matrix elements required are given in table XX {[}list for X, A and B states\ldots

% Although the retrieved coefficients $C_{K,Q}^{L,M}$ contain information on the ionization dynamics, it is preferable to use the resulting ADMs within the full ionization framework of Eq.~\ref{eq:beta} to directly retrieve the intrinsic terms $\boldsymbol{D}_{hl}^{\Gamma\mu}(q)$. 
% It is of note that, even in the sanitised form shown here, this formalism is complicated, and includes interferences between different partial waves, polarization geometries and ionization channels.
Although the retrieved coefficients $C_{K,Q}^{L,M}$ contain information on the ionization dynamics \cite{Makhija2016}, it is preferable to use the resulting ADMs within the full ionization framework of Eq.~\ref{eq:beta} to directly retrieve the intrinsic molecular ionization dynamics defined by $\boldsymbol{D}_{\alpha}$. This constitutes the second stage of the bootstrapping procedure. The multiple interfering pathways set the requirement for a high information-content dataset, and the equations are highly non-linear multivariate quadratic equations. % \textcolor{red}{[Varun's 'maths refs' here]}. 
Careful sampling of the solution hyperspace is required in order to ensure unique results: this step was therefore broken down into sub-steps to allow (a) statistical sampling and (b) further bootstrapping. In (a) repeated coarse fits with minimal information content (selected experimental measurements with distinct ADMs) and randomised seed parameters were employed \cite{Hockett_2015a}, while (b) took the best fit result(s) as seeds for fits with higher information content and/or more stringent convergence criteria, based on computational time and desired precision. % (see SM).

Fig. \ref{fig:BLMfits} illustrates this method and example results. 
% The main panels show $\beta_{0,0}(t)$ traces for H5 ionization, with full angle-resolved interferograms $S(\theta,t)$ shown for a few time-steps. An example of the complete $\beta_{L,0}(t)$ traces are also shown for the $X$% -channel. 
% The wealth of data, and the high information content of $S(\theta,t)$, is clear. % The interferograms $S(\theta,t)$ 
The $X$-channel is particularly sensitive to the axis distribution, with significant changes in both the yields and angular distributions. %, while the $A$-channel is much less sensitive and shows less dramatic variations in $S(\theta,t)$. 
% In the data, $L_{max}=6$ is observed, suggesting that the lesser of $K_{max}$ or $l_{max}$=3; based on this observation, and the fact that $K_{max}=6$ from the linear fitting stage, a cut-off of $l_{max}=4$ was assumed in the matrix element retrieval.
% For the $X$-channel, 
In this case, two unique fit results (sets of matrix elements $\boldsymbol{D}_{\alpha}$) were obtained via statistical sampling at a coarse level (11 temporal points over the revival features), and fine-tuning via the bootstrapping methodology outlined above (finally incorporating 89 temporal points), led to a single best-fit solution (6 complex-valued matrix elements, results as shown in Fig. \ref{fig:BLMfits}(a)). For the $A$ and $B$-channels the data becomes increasingly noisy as the yields decrease, and also indicates much less dramatic dependence on the axis distribution. In these cases, adequate fits were obtained at the coarse level (as indicated in fig. \ref{fig:BLMfits}(b)), %, although for the $A$-channel a unique parameter set was not obtained; 
but further bootstrapping was not explored in either case. For the $A$-channel, three best fit parameter sets were obtained (7 matrix elements), differing only in the perpendicular continuum waves. % (see Fig. \ref{fig:MFrecon}).
For the $B$-channel data % (see SM)
, a single best fit parameter set was obtained (5 matrix elements), although four additional parameter sets were within 1\% of the best fit (defined by the minimal residual, $\chi^2$ \cite{Bevington1992}), and differed in the parallel continuum. In these cases, additional data and/or cross-checks on the determined matrix elements are therefore desirable to confirm their validity. The full set of fit results and matrix elements determined (including the associated uncertainties) are given in the SM.

\begin{figure}[h!]
\begin{center}
\includegraphics[width=1\columnwidth]{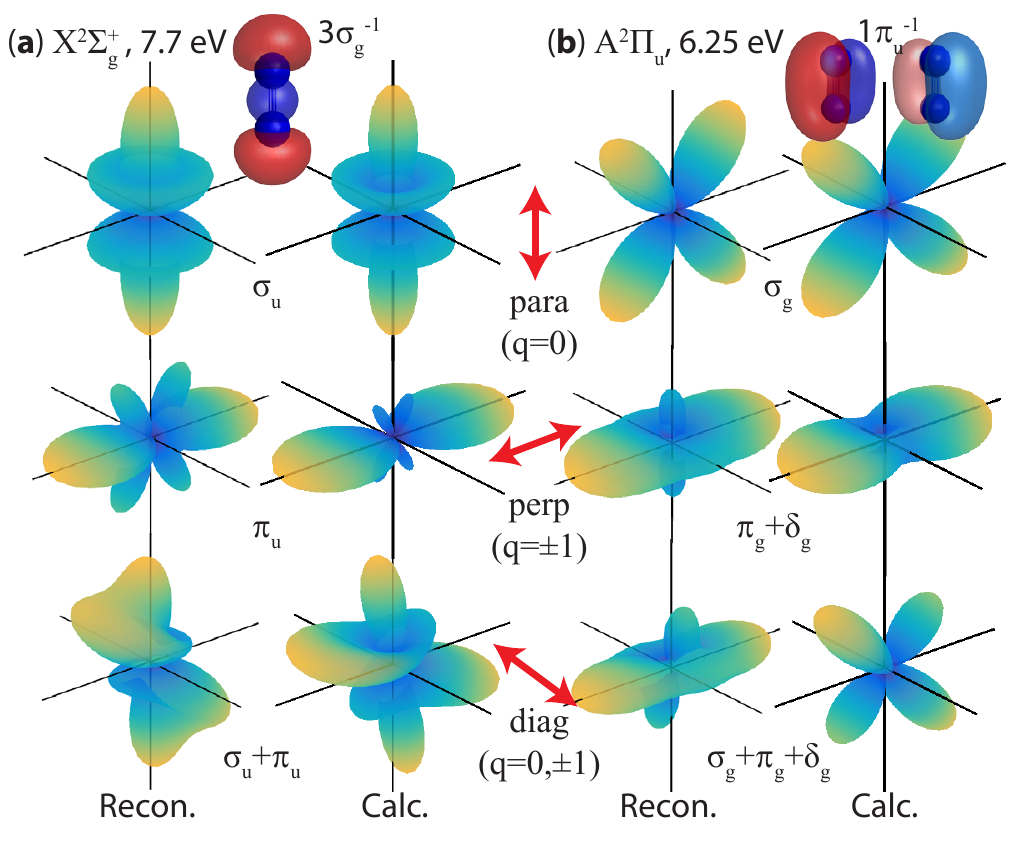}
\caption{{\label{fig:MFrecon}Molecular frame reconstruction as a function of ionizing orbital and polarization geometry. (Recon) MFPADs reconstructed from the experimental analysis; (Calc.) determined from \emph{ab initio} calculations. Each panel also shows the ionizing orbital, and labels indicate the MF polarization geometry (photon projection $q$) and corresponding symmetries for the ionization continua accessed. % An extended presentation is given in the SM.%
}}
\end{center}
\end{figure}

Finally, from the matrix elements obtained, the molecular frame (MF) photoelectron interferograms may be calculated, and compared with \emph{ab initio} results. In this case, the interferograms are no longer restricted by the LF symmetry, hence this provides a sensitive test of the retrieved matrix elements. Fig. \ref{fig:MFrecon} shows these results, and comparison with \emph{ab initio} results obtained using ePolyScat \cite{Gianturco_1994,Natalense_1999}. % The calculations employ a single active electron (SAE) single channel treatment, and the dipole approximation. 
Results are shown for three different polarization geometries in the MF. The parallel and perpendicular cases cleanly separate sets of matrix elements $\boldsymbol{D}_{\alpha}$ by symmetry, % with photon projections in the MF $q=0$ and $q=\pm 1$ respectively, 
while the diagonal result mixes these components and provides a particularly stringent test of the results \cite{Lucchese_2002}. Compared to the LF, the MF results show a wealth of structure. The experimentally reconstructed and computed results for the $X$-channel are in good agreement, while for the $A$ channel the agreement is variable, % and both the experimental and computed results are expected to be less accurate.
 consistent with larger uncertainties in the retrieved matrix elements \cite{SM}. % The interferograms for the diagonal case are a particularly stringent test of the results, since they are sensitive to both the relative magnitudes and phases of the parallel and perpendicular continua, and these phase relations are non-trivial to obtain \cite{Lucchese_2002}. 
% Again, the detailed structures obtained here, and the agreement with computational results, suggests that the matrix elements have been accurately determined. 
Photoionization of $N_2$ has attracted much interest due to the $\sigma$ continuum shape resonance \cite{Lucchese_1986}. Significantly, the MF results (Fig. \ref{fig:MFrecon}) for the $X$ and $B$ $\sigma$ continua are similar to direct MF measurements in the same photoelectron energy region presented in refs. \cite{Shigemasa_1995,Yagishita_2005} for 2 and 9~eV photoelectrons; while those studies investigated dissociative core ionization, the scattering and shape-resonance behaviour of the $\sigma$ continuum accessed is analogous.

From the \emph{ab initio} perspective, photoionization calculations present a formidable challenge and there are a dearth of MF measurements for comparison and validation, due to the experimental difficulty of such measurements; thus the MF results herein can alternatively be viewed as an excellent test for theory \cite{Lucchese_1986,Semenov_2000,Lin_2002, Lucchese_2002,Jin_2010}. From this perspective, consistency between the \emph{ab initio} and experimental reconstructions can be taken as a good indicator that both methodologies are robust. %, and a means to test for the breakdown of the approximations.

In this work, time-domain measurements, utilizing molecular alignment techniques, have been demonstrated as a means to ``complete" photoionization studies. The time-domain data provided sufficiently high information content to reliably extract the axis distributions and matrix elements for three different ionization channels. A bootstrapping fitting methodology provided a flexible and robust analysis route, and MF reconstructions provided an additional stringent test of the physical parameters so determined, and an illustration of their predictive power. The main advantage of this methodology is that it is completely general in principle (although may be restricted by symmetry in certain cases \cite{SM}), and applicable to any molecular ionization problem, provided that the preparation and propagation of the rotational wavepacket remains decoupled from other molecular dynamics. All relative magnitudes and phases of the ionization matrix elements can be determined and % including their signs \cite{Suzuki_2012}, in contrast to most other methods; % [NOT sure if this is strictly true - NOPE, Suzuki relies on energy mapping & comparison with theory here]
MF interferograms can be obtained without experimentally challenging MF measurements, which require molecular orientation or fragmentation. % The fitting method outlined allows for a balance between accuracy and computational time, according to the problem and data at hand. 
The disadvantage is that this fitting methodology is not a black-box procedure, and requires detailed analysis (see SM). While this methodology has been demonstrated herein for the simplest case of a homonuclear linear molecule, the theory and analysis protocol allows for arbitrary asymmetric molecules, although such cases will clearly be more challenging \cite{Makhija2016a}. Some improvements may be possible here, for example the use of other fitting algorithms (genetic algorithms and homotopy methods \cite{Sommese2005}), which may allow for a more robust approach less sensitive to local minima in the solution hyperspace, and the use of GPUs to massively parallelise the computations. %, since the computations at each $t$ are independent. 

Experimentally, the use of an extended harmonic spectrum would allow for the observation and analysis of many more photoelectron bands \cite{Rouz_e_2012}, thus providing a route to obtaining even higher degrees of multiplexing in the measurements, and MF reconstruction for a range of channels and energies from a single time-domain experimental dataset. In this light, the generality of the method suggests such measurements, combined with high-repetition rate laser sources, provide a route to highly multiplexed ``complete" experiments and MF reconstruction for a range of molecules, and a tractable method for high-resolution dynamical MF imaging \cite{hockettBrief2014}, which is otherwise typically restricted by experimental prerequisites and data-acquisition time-scales \cite{Hockett_2013}.

% (ionization channel and photoelectron energy) measurements and MF reconstruction from a single time-domain experimental dataset.

% Overall, the use of rotational wavepackets is of clear utility for complete photoionization experiments, and molecular frame reconstruction.

Acknowledgements: We thank the US Army Research Office, for funding under contract  W911NF-14-1-0383. AS thanks the NSERC Discovery program for financial support. 

Supplementary Material  \cite{SM} available at \href{https://dx.doi.org/10.6084/m9.figshare.4480349}{DOI: 10.6084/m9.figshare.4480349}.

\bibliographystyle{apsrev4-1}
% \bibliography{bibliography/converted_to_latex}
\bibliography{converted_to_latex}

\end{document}